\documentclass[prb,preprint,amsmath,amssymb]{revtex4-1}
\usepackage{graphicx}
\usepackage{epsfig}
\usepackage{dcolumn}
\usepackage{bm}
\begin{document}
\def\Tau{\mbox{\boldmath $\tau$}}

\title{\bf Magneto- and Baro- Caloric Responses in Magnetovolumic Systems}

\author{Eduardo Mendive-Tapia}
\affiliation{Department of Physics, University of Warwick, Coventry CV4 7AL, U.K.}
\author{Teresa Cast\'an}
\affiliation{Departament d'Estructura i Constituents de  la
  Mat\`eria. Facultat de  F\'\i sica.  Universitat de
  Barcelona. Diagonal, 647, E-08028 Barcelona, Catalonia.}

\date{\today}
\begin{abstract}
By means of a mean-field model extended to include
magnetovolumic effects we study the effect of external fields  on the
thermal response characterized either by the isothermal entropy change
and/or the adiabatic temperature change. The model includes two
different situations induced by the magnetovolumic coupling. (i) A
first order para- ferromagnetic phase transition  that entails a
volume change. (ii) An inversion of the effective exchange interaction
that promotes the occurence of  an  antiferromagnetic phase  at low
temperatures. In both cases, we study  the  magneto- and baro-caloric
effects as well as the corresponding cross caloric responses. By
comparing the present theoretical results with avaliable experimental
data for several materials we conclude that the present
thermodynamical model  reproduces the general trends associated with
the considered caloric and cross caloric responses.
\end{abstract}

\maketitle

\section{Introduction}
\label{Intro}
Solid state refrigeration based on caloric effects is currently a very
active research topic because of the possibility of developing new
friendly alternative refrigeration devices
\cite{Sandeman2012}. Caloric effects originate from the thermal
response of every thermodynamic system to changes induced by the
variation (either application or removal) of an external field
\cite{Manosa2013}.  Depending on the external field, the corresponding
caloric effect  is called magnetocaloric (magnetic field)
\cite{Krenke,sandeman1,Oliveira2010,Planes2009,Planes2014}, barocaloric
(hydrostatic
pressure)\cite{Manosa2010,Manosa2011,Oliveira2011,Stern2014},
electrocaloric (electric
field)\cite{Neese2008,Lu2011,Moya2013,Lisenkov2013}, elastocaloric
(mechanical stress)\cite{Bonnot2008,Xiao2013,Nikitin1992}, and
toroidocaloric (toroidic field)\cite{Castan2012}. The two limiting
situations correspond to either varying the external field
isothermically or adiabatically. In the first case a change in the
entropy is induced while in the second the system responds with a
temperature shift. These isothermal change of entropy and adiabatic
change of temperature are commonly used in order to quantify the
caloric response of a given system. The interest is to be able to
induce a large caloric effect in response to small or moderate
variations of the external field.  Indeed, this is most likely to
occur in the vicinity of a phase
transition\cite{Manosa2010}. Moreover,  systems with coupled degrees
of freedom might respond to different species of external fields.
This gives rise to the so called field-tune caloric effect and
multicaloric effect \cite{Fahler2011,Vopson2012,Meng2013,Moya2014}.
In the first situation, the secondary field is kept constant during
the variation of the primary field. While the primary field
effectively drives the caloric response, the secondary field  allows to
adjust the best operative conditions. In the second situation, the
multicaloric effect refers to  the variation of two or more fields
either simultaneously or sequentially. For instance, in the case of
systems with magnetoelastic coupling, the interplay between magnetism
and  elastic properties allows to induce a caloric response in the
system by the application of either a magnetic field  or/and a
mechanical field (hydrostatic pressure or stress). In the present
investigation we shall  focus in magnetovolumic systems. 

 Magnetovolumic effects arise as a special case of  magnetoelastic
 coupling in which variations in the magnetization are accompanied by
 an isotropic change in volume. Such variations may
 be spontaneous, through a phase transition, or forced by the
 application of an external field.  The interaction between volume and
 magnetism results in  the interrelation between  magneto- and baro-
 caloric effects observed experimentally in different materials
 \cite{Fujita12003,Fujita22003,Lyubina2008,Annaorazov1996,Annaorazov2002}. 

The present theoretical study is based on a mean-field Ising
model\cite{Ising} for phase transitions extended to include coupling
between volume and magnetism. Interestingly, the model allows to study
two different situations. In the first one, the magnetovolumic
coupling induces a first-order para-ferromagnetic phase transition
that can be modified by the application of either a hydrostatic
pressure or/and a magnetic field. In the second situation, the
interplay between volume and magnetism originates a strong first-order
antiferro-ferromagnetic transition that is responsive to the
application of both hydrostatic
pressure and magnetic field. Effective and mean-field
approaches\cite{trigero1,Bean1962,Yamada2001,Ranke2004,Ranke2006,Ranke2009,Valiev2009,menyuk1}
have been used previously to investigate magnetovolumic
effects. Compared with these prior investigations, the present work
incorporates the occurrence of a metamagnetic transition and the study
of caloric  and cross-caloric effects. 

The paper is organized as follows. In section \ref{Model} we briefly
resume the main aspects of the model  and the thermodynamics of
caloric effects. In section \ref{Ferro} and \ref{Inversion} we solve
numerically the model  with special attention to the 
metamagnetic transition (section IV). We first obtain the phase diagram
and  study how the different transition temperatures change with applied
fields (either hydrostatic pressure and/or magnetic field) and next we present the results for both the baro- and  magneto- caloric effects. 
In section \ref{Discussion} we compare our results 
with experimental data available for magnetic and metamagnetic materials. We finally outline our main conclusions in section \ref{conclusions}.
 
\section{Modeling and thermodynamics of caloric effects}
\label{Model}
The model under consideration is based on the statistico-mechanical
mean-field Ising model extended to include magnetovolumic effects. The
starting point is a free-energy, consisting of the sum of two
contributions,  $f=f_M+f_C$.  The first contribution, $f_M$, that accounts for the
magnetic degrees of freedom, can be expressed in terms of both the
ferromagnetic ($m$) and the antiferromagnetic ($x$) order parameters
simultaneously \cite{Ising}
\begin{equation}
\begin{split}
 & f_{M} (T,m,x)  =    -\frac{Jz}{2}(m^2-x^2) - k_B T \ln2+
  \frac{k_B T}{4}  [ (1+m+x) \ln (1+m+x) +  \\ &    +  (1+m-x) \ln
    (1+m-x)     +(1-m+x) \ln (1-m+x) + \\  & +  (1-m-x) \ln
    (1-m-x) ],  
\label{EQ1}
\end{split}
\end{equation}
Hereafter the exchange interaction is fixed to be positive ($J>0$). In
that case, the previous free energy (\ref{EQ1}) produces a
continuous para-ferromagnetic  phase transition at $T_c=zJ/k_B$,  being $z$ the
number of nearest neighbours and $k_{B}$  the Botzmann constant.  The
second contribution, $f_C$, incorporates the magnetovolumic coupling
and includes magnetostriction coupling of both order parameters, $m$
and $x$, to the relative volume change $w=\frac{\delta
  \Omega}{\Omega}$, where $\Omega$ is some reference
volume. Restricting the coupling terms to the  minimum order allowed
by symmetry, one may write: 
\begin{equation}
f_C(m,x,w)= \frac{\alpha_0}{2}w^2 - (\alpha_1 m^2 + \alpha_2
x^2)\frac{w}{2}.  
\label{EQ2}
\end{equation}
We have also included a purely elastic contribution, with
$\alpha_0$ being proportional to the inverse of the compressibility.  Furthermore, in order
to account for pressure effects as well as for the interplay with  an external
magnetic field, we introduce the following Legendre transform
to the total free-energy:
\begin{equation}  
g= f(T,m,x,w)-Hm+ \Omega P w,
\label{EQ3}
\end{equation}
where $\mathnormal {g}$ stands for the Gibbs free-energy, $P$ is the
hydrostatic pressure and $H$ is the external magnetic field.  In
expression (\ref{EQ2}) $\alpha_1$ is the magnetostriction coefficient
that gives rise to a first-order phase transition from a paramagnetic
($\cal{P}$) phase to a ferromagnetic ($\cal{F}$) phase when lowering
the temperature. The coefficient $\alpha_2$ causes an inversion-exchange 
of the effective interaction so that an
antiferromagnetic ($\cal{AF}$) order might exist for some range of
model parameters and  applied external fields.

We remark that the Landau-based phenomenological expansion in eq. (\ref{EQ2}) is based
on symmetry considerations and that it intends to describe the effects
of the  interplay between volume and magnetism rather than to address its
physical origin. The physical mechanism that originates such
interplay and the way it operates can be different from one system to
another. Nevertheless, the symmetry-based coupling in (\ref{EQ2}) is
present in all magnetic materials although in some cases can be
negligible.  Moreover, coupling coefficients  are material dependent
and can be functions of  chemical composition and  valence electron
concentration, among others.  It is worth mentioning that the
linear-quadratic coupling between volume change and magnetization has
been used previously through a prescribed linear dependence of
the Curie temperature with the volume change\cite{Bean1962}.

It is convenient to get rid of the (secondary) order parameter $w$ by minimizing
expression (\ref{EQ3}) with respect to $w$. One gets:
\begin{equation}
w=\frac{1}{2\alpha_0} \left [(\alpha_1 m^2 + \alpha_2 x^2)- 2P \Omega \right ]
\label{EQ4}
\end{equation}
This constitutive equation verifies the following Maxwell relation \cite{Planes20142,Vopson2012}, 
\begin{equation}
\left (\frac{\partial m}{\partial P} \right )_{T,H} = - \Omega \left
(\frac{\partial \omega }{\partial H} \right )_{T,P},
\label{EQ7bis}
\end{equation}
that underlines the origin of the multicaloric response.
Therefore, the Gibbs free-energy per magnetic particle, in reduced units, along the optimum
path involving $m$, $x$, and $w$ given by (\ref{EQ4}), is:
\begin{equation}
\begin{split}
&  g^* = \frac{g}{zJ}= -\frac{1}{2}(m^2-x^2) - T^* \ln2 +
  \frac{T^*}{4}  [ (1+m+x) \ln (1+m+x) +   \\ &  +  (1+m-x) \ln
    (1+m-x)  +(1-m+x) \ln (1-m+x)   +  \\ & + (1-m-x) \ln
    (1-m-x) ]- \frac{1}{8 \alpha_0^*}[(\alpha_1^* m^2+\alpha_2^* x^2)- 2P
      \Omega^*]^2- H^* m.
\label{EQ5}
\end{split}
\end{equation}
Where the superscript ($*$) indicates that the magnitude is normalized to $zJ$. 
We take  $\alpha_0^*=1$, $\Omega^*=1$, without loss of generality.

When a given external field ($Y$) is modified (applied/removed)
isothermally, the corresponding caloric effect is related to the
entropy change of the system that can be obtained from fundamental
Thermodynamics \cite{Planes2009,Planes20142}. Indeed, for a
finite change of the field ($Y=0 \,\rightarrow \,Y \neq 0$), the
corresponding field-induced isothermal entropy change will be given
by:
\begin{equation}
 \Delta S(T,0 \rightarrow Y) = S(T,Y)-S(T,0)  =\int_0^Y \left(
 \frac{\partial S}{\partial Y} \right )_T dY = \int_0^Y \left(
 \frac{\partial X}{\partial T} \right )_Y dY,
\label{EQ6}
\end {equation}
where we have used the appropriate Maxwell relation and $X$ is the
thermodynamically conjugated variable to the field $Y$. The present
model can be applied to study both magnetocaloric (MCE) and
barocaloric (BCE) effects corresponding to ($Y=H$, $X=m$) and 
($Y=-P$, $X=w$) respectively. Indeed, the entropy can be
directly obtained from (\ref{EQ5}) by taking into account that,
\begin{equation}
\begin{split}
& S(m,x) = - \left [\frac{\partial g^{*}}{\partial T^{*}} \right ]_{H,P}  =
   \\ &  \ln 2 - \frac{1}{4} [ (1+m+x) \ln (1+m+x) + (1-m+x) \ln
    (1-m+x)    + \\ & + (1+m-x) \ln (1+m-x)+(1-m-x) \ln (1-m-x) ]
\end{split}
\label{EQ7}
\end{equation}
In the expression above, $m=m(T^{*},H^{*},P)$ and $x=x(T^{*},H^{*},P)$ are the
equilibrium order parameters obtained after minimization of the
free-energy (\ref{EQ5}). In addition, for a given caloric effect, the entropy change
 should depend on the
(secondary) tunning  field.  For instance, the pressure-tune  MCE at a
given constant value of $P$,  is characterized by the entropy difference
$\Delta S(T,0 \rightarrow H,P) =
S(T,H,P)-S(T,0,P)$. Alternatively, the magnetic
field tuned BCE depends on the value of the (secondary) applied
magnetic field $H$ and it is given by $\Delta S(T,H,0
\rightarrow P) = S(T,H,P)-S(T,H,0)$. Notice that
by tunning the secondary field, it is possible to
adjust the most optimum temperature range for the caloric
effect.  Moreover, in the case of the multicaloric effect   the
corresponding entropy change is given by $\Delta S(T,0\rightarrow
H,0\rightarrow P)= S(T,H,P)-S(T,0,0)$ and both,
pressure and magnetic field, are applied/removed  simultaneously (or
sequentially).  Given that the entropy is a state function that
depends only on the current state of the system,  it is easy to show
that \cite{Planes2014}
\begin{equation}
\begin{split}
& \Delta S(T,0 \rightarrow H,0 \rightarrow P)  = \\ & \Delta S_{MCE} (T,0
  \rightarrow H, 0)+ \Delta  S_{H-BCE} (T,H, 0 \rightarrow P) = \\ &\Delta
  S_{BCE} (T,0,0 \rightarrow P)+ \Delta  S_{P-MCE} (T,0 \rightarrow H, P), 
\end{split}
\label{EQ8}
\end{equation}
where $\Delta S_{MCE}$ stands for MCE, $\Delta S_{P-MCE}$ for P-tune MCE, 
$\Delta S_{BCE}$  for BCE and  $\Delta S_{H-BCE}$ for H-tune BCE. 
For the sake of clarity we shall keep this notation along the present work.

When the  external field is changed adiabatically,
 the subsequent temperature change can be expressed as
\begin{equation}
 \Delta T(0\rightarrow\,Y) = -\int_{0}^{Y}{\frac{T}{C}\left(
   \frac{\partial S}{\partial Y}\right)_{T} dY }=
 -\int_{0}^{Y}{\frac{T}{C}\left( \frac{\partial X}{\partial
     T}\right)_{Y} dY },
\label{EQT9}
\end{equation}
where again we have used the appropriate Maxwell relation. $C$ is the
heat capacity and the external field is varied from $Y=0$ to $Y\neq
0$.  Note that the previous thermodynamic expression (\ref{EQT9})
involves the total entropy of the system.  Nevertheless, the model
entropy in Eq. (\ref{EQ7}) only accounts for the magnetic contribution. 
Consequently, such entropy returns values for the calculated adiabatic
temperature variations (\ref{EQT9}) definitively unphysical.  To improve 
this, we consider the lattice contribution per particle in the Debye approximation,
given by \cite{Oliveira2010}
\begin{equation}
S_{v} =
    k_B \Biggl[-3\ln\left(1-e^{-\frac{T}{\theta_{D}}}\right)
    +12\left(\frac{T}{\theta_{D}}\right)^{3}\int_{0}^{\theta_{D}/T}\frac{x^{3}}{e^{x}-1}dx\Biggr],
\label{EQTS}
\end{equation}
$\theta_{D}$ being the Debye temperature.  We now proceed by merely appending
expression (\ref{EQTS}) to the magnetic entropy
(\ref{EQ7}). Physically, this additional term plays the role of a
thermal bath (or reservoir) replacing the effects of the remaining
degrees of freedom not considered explicitly in the model.  This is a
quite usual approach in Statistical Mechanics. We notice that eventual
influences due to a volume dependence of the lattice  entropy
contribution or electronic effects are not considered explicitly
here. Nevertheless, when looking at the behaviour of a given specific
material, such effects  can be relevant and therefore should be taken
into account.  
\section{Field-induced Ferromagnetic Transition}
\label{Ferro}
\begin{figure}[ht]
\centering
\includegraphics[clip,scale=0.5]{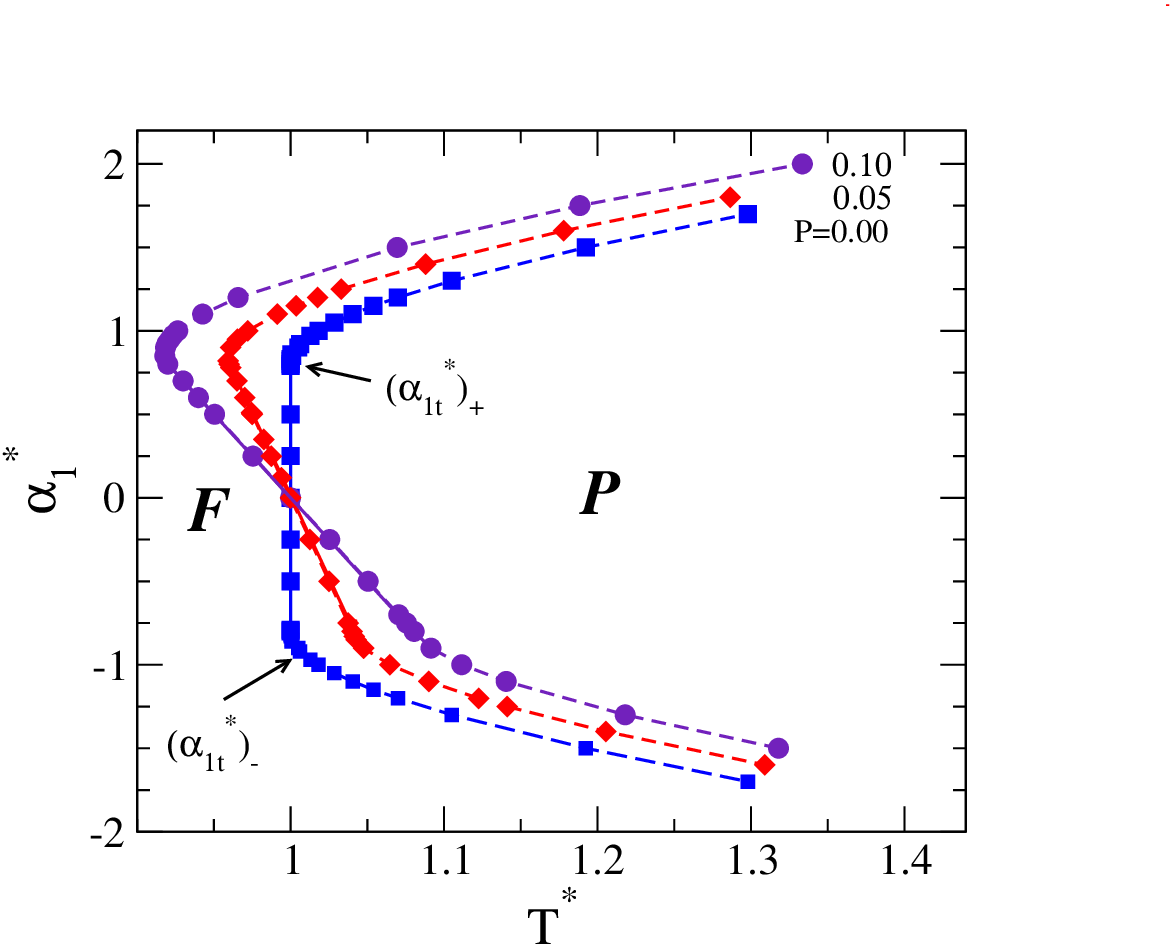}
\caption{(Color online) Transition temperature versus the coupling parameter  $\alpha_1^*$
  for $H^{*}=0$.  The three curves correspond to selected
  values of the pressure $P$ as indicated. Second order transitions
  are denoted by  solid lines whereas  first order transitions
  by   dashed lines. Both curves intersect at the two tricritial
  points $\left(\alpha_t^*\right)_{+}$ and $\left(\alpha_t^*\right)_{-}$.}
\label{FIG1}
\end{figure}

In this section we briefly summarize the main results obtained by
solving numerically the minimal model that allows for a  discontinuous
$\cal{P}$-to-$\cal{F}$-phase transition, involving volume variation,
under the application of an external field (either $P$ or/and $H$).
This corresponds to set $\alpha_2^*$=0 in  eq.(\ref{EQ5}), obtaining
the following  free energy function:
\begin{equation}
\begin{split}
&  g^* = -\frac{m^2}{2} - T^* \ln 2  + \frac{T^*}{2} [ (1+m)
    \ln (1+m) + \\ & + (1-m) \ln (1-m) ] - \frac{1}{8
    \alpha_0^*}(\alpha_1^* m^2 + 2 \Omega^* P)^2  - H^* m,
\label{EQ9}
\end{split}
\end{equation}
where the ${\cal AF}$ order parameter is  $x=0$ for all range of $T^*$
and $\alpha_1^*$.  For given values of the external fields, a further
direct numerical minimization of (\ref{EQ9}) with respect to $m$
renders the thermodynamical solutions for $m$($T^*$,$H^*$,$P$).
Afterwards, it is possible to compute all  thermodynamic quantities of
interest.  In the present study we restrict ourselves to some
representative results.  Firstly, figure \ref{FIG1} shows the phase
diagram as a function of the coupling parameter  $\alpha_1^*$, for $
H^*$=0 and for three different values of the pressure $P$=0, 0.05,
0.1, as indicated. Each curve exhibits  two tricritical points
($\left(\alpha_{1t}^*\right)_{\pm}$, $T^{*}_{t}$) that change with the
external pressure $P$.  For $\left(\alpha_{1t}^*\right)_{-} <
\alpha_1^*< \left(\alpha_{1t}^*\right)_{+}$ the transition is
continuous whereas for $\alpha_1^*>\left(\alpha_{1t}^*\right)_{+}$ and
$\alpha_1^*<\left(\alpha_{1t}^*\right)_{-}$ it is discontinuous.  The
Curie temperature $T_c^*$ for the continuous transition \footnote{It
  can be obtained from a Landau expansion of the free energy
  (\ref{EQ9}) and then require that the harmonic coefficient be equal
  to zero} is  given by $T_c^*(P^*) = 1 -\frac{\alpha_1^*}{\alpha_0^*}
\Omega P$.  An inspection of Fig.\ref{FIG1} reveals that for $P=0$ the
sign of $\alpha_1^*$ is irrelevant  whereas under the application of
an external pressure,  $T_c^*$ (continuous line)  may decrease or
increase with $P$, depending on whether the coupling parameter
$\alpha_1^*$ is positive or negative respectively.  Beyond the
tricritical points, the transition temperature for the  first order
transition (dashed line)  increases with $\alpha_1^*$, regardless its 
sign. Below, we summarize the main
results obtained for the MCE and BCE behaviours for a representative
value of $\alpha_1^*= \pm$ 1.10, for which the transition is
discontinuous.  

\begin{figure}[ht]
\begin{center}
\includegraphics[clip,scale=0.5]{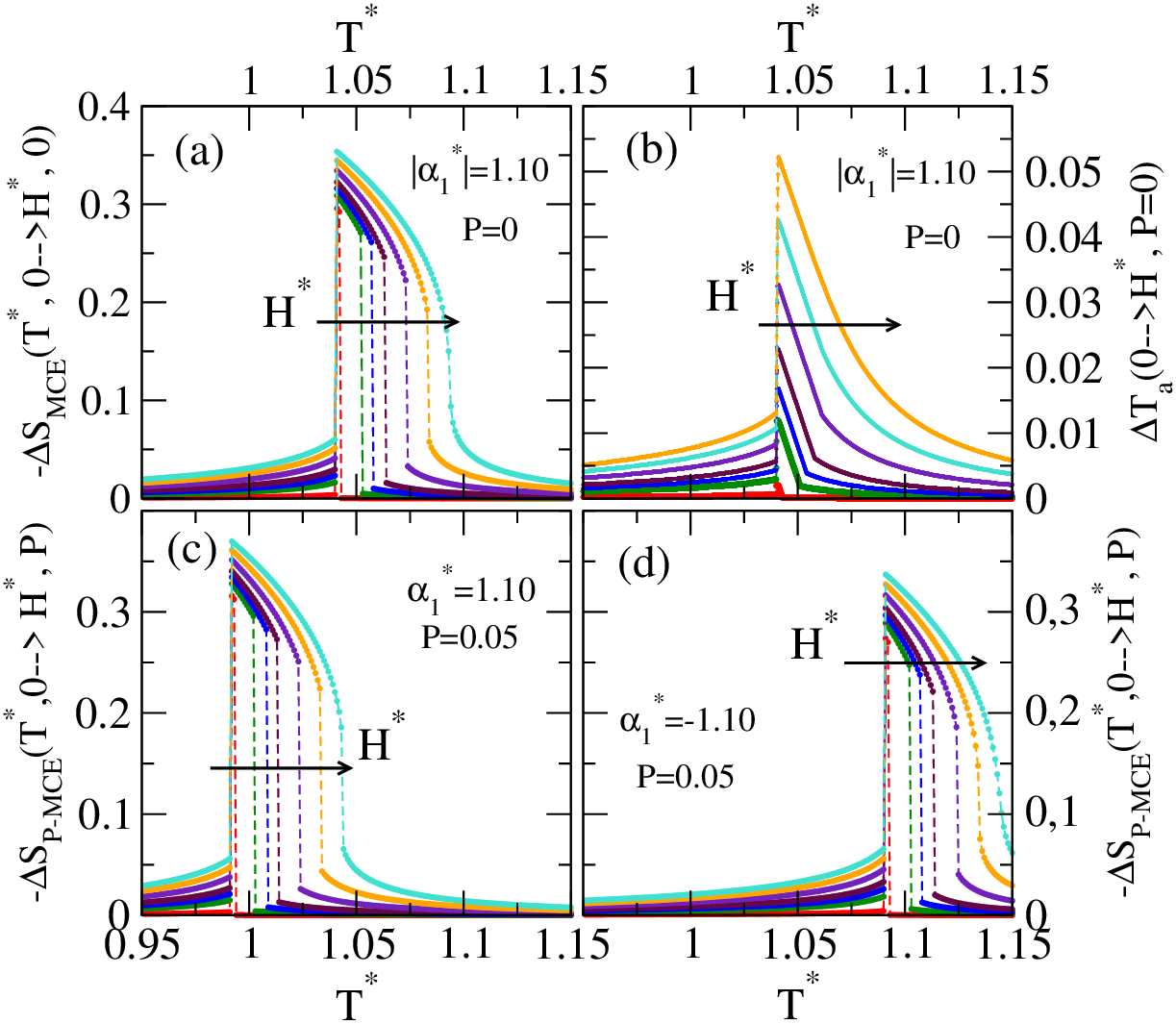}
\end{center}
\caption{(color online) MCE under the application of increasing values
  of the magnetic field $H^*$ (denoted by an arrow). (a) Isothermal
  entropy change at $P=0$, (b) adiabatic temperature change at $P=0$,
  (c) and (d) isothermal entropy change at $P=0.05$ for
  $\alpha_1^*=1.10$ and  $\alpha_1^*=-1.10$ respectively.}
\label{FIG2}
\end{figure}

Figure \ref{FIG2} displays the temperature behavior of the MCE for
different external fields. In the upper panels  we have plotted  (a)
the isothermal entropy change  $\Delta S_{MCE}(T,0\rightarrow H^*,
P=0)$  and (b) the adiabatical temperature shift  $\Delta
T_{a}^*(0\rightarrow H^*, P=0)$, for increasing values of the applied
magnetic field (denoted by an arrow).  Both behaviours are consistent
with a conventional MCE.  To illustrate the behavior of the external
pressure on the MCE,  we have plotted the $\Delta S_{P-MCE}$ at
$P=0.05$ for $\alpha_1^*=1.1$ (c) and $\alpha_1^*=-1.1$ (d). The
effect of $P$ is to shift the MCE peak either to lower (c) or higher
(d) temperatures depending to sign of $\alpha_1^*$, accordingly to the
tendency of promoting the phase with lower volume. 
\begin{figure}[ht]
\centering
\includegraphics[clip,scale=0.5]{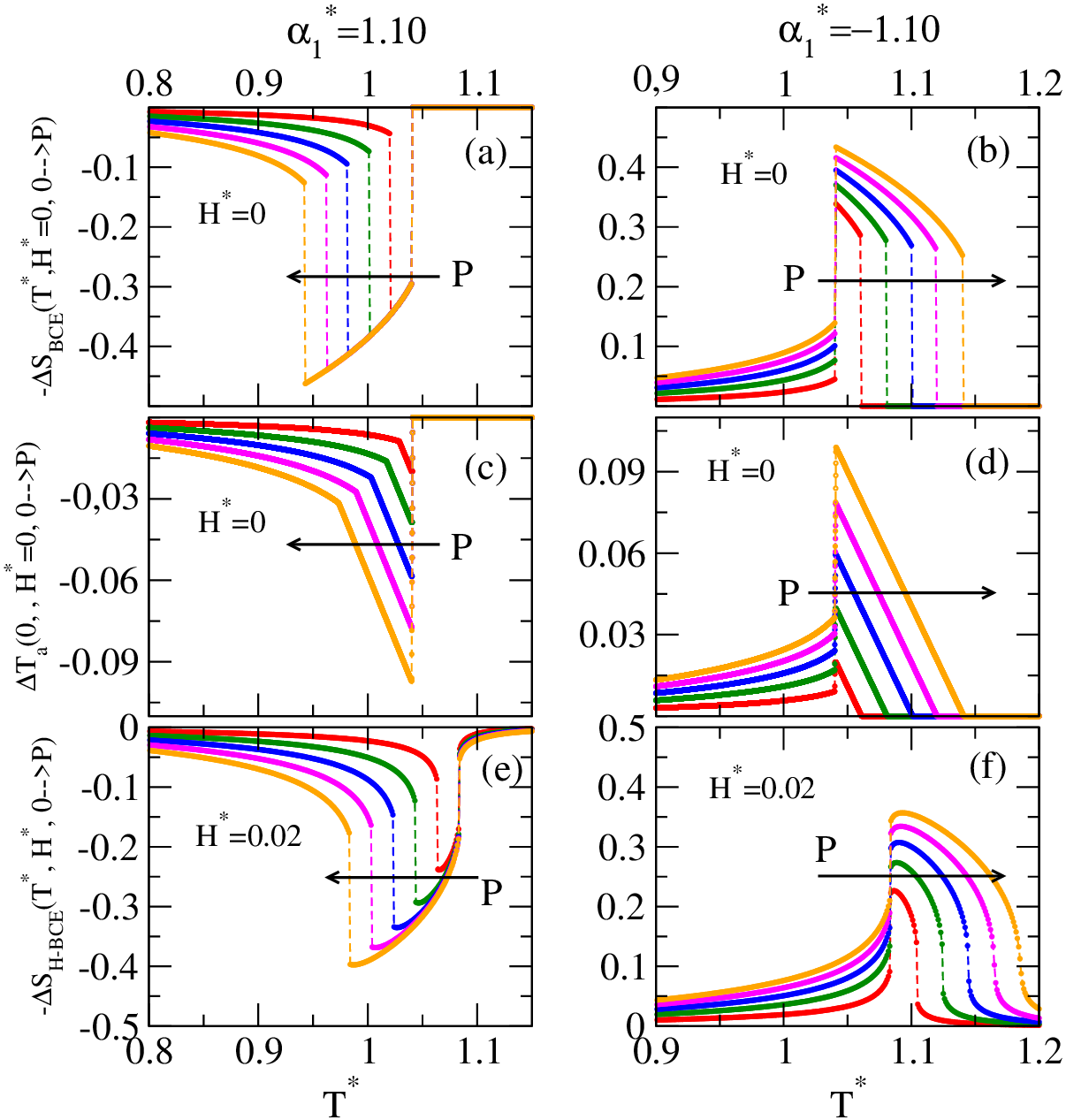}
\caption{(Color online) BCE for   two representative values of
  $\alpha_1^*=1.10$ (left column) and $\alpha_1^*=-1.10$ (right
  column) for increasing values of the applied pressure $P^*$ . Panels (a)
  and (b) display the isothermal entropy change at $H^*=0$, (c) and (d)
  the corresponding adiabatic temperature shift and (e) and (f) the
  isothermal entropy change at $H^*=0.02$.} 
\label{FIG3}
\end{figure}

The results for the BCE are  shown in Fig. \ref{FIG3}.  As expected,
one obtains different behaviors depending on the sign of $\alpha_1^*$.
Essentially, for $\alpha_1^*$=1.10 the BCE is inverse whereas for
$\alpha_1^*$=-1.10 it is conventional. Consequently, the entropy
increases (a) or decreases (b) when the pressure is applied
isothermally. Likewise, the system cools down (c) or warms up (d)
when the pressure is applied adiabatically.  The effect of $H^*$ on
the BCE is shown in the lower panels of the same figure for
$\alpha_1^*=1$ (e) and $\alpha_1^*=-1$ (f). As can be observed, the
application of the secondary field  $H^*$ shifts the caloric response
towards higher temperatures and reduces the peak, regardless of the
sign of $\alpha_1^*$. This reflects the natural tendency of the
external $H^*$ to  promote the (ordered) $\cal{F}$-phase with lower
entropy.   In summary,  the effect of increasing $H^*$ on the BCE is
to attain higher temperatures,  at expenses of reducing the caloric
response.

\section{The Metamagnetic Transition}
\label{Inversion}
\begin{figure}[ht]
\centering
\includegraphics[clip, scale=0.5]{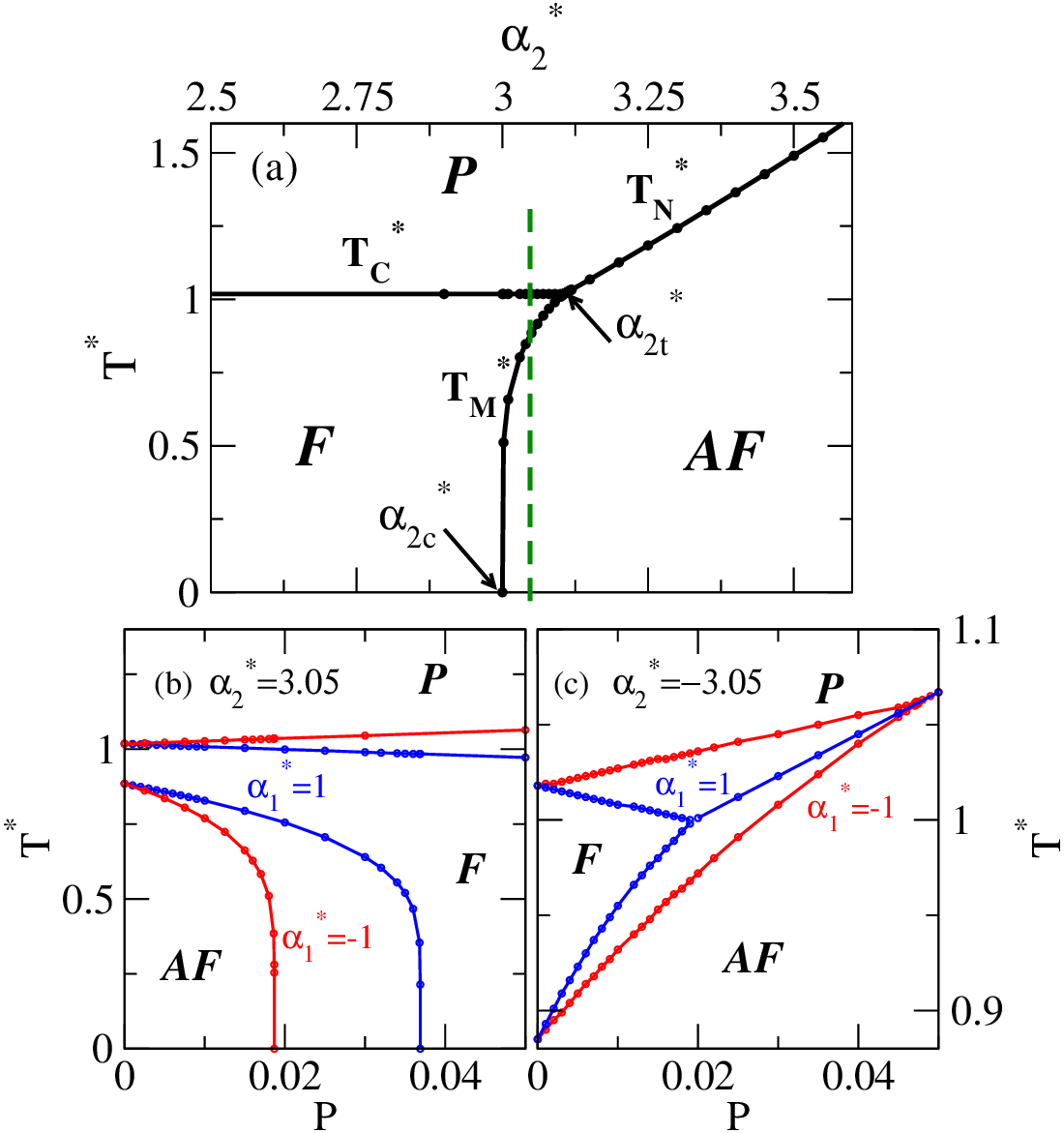}
\caption{(Color online) (a) Phase diagram for the exchange-inversion
  model in the region of positive $\alpha_{2}^*$ at $H^{*}=0$, $P=0$ and
  $\alpha_{1}^*=1$.  The green dashed line denotes the value of
  $\alpha_{2}^*=3.05$ set for the present calculations.  Lower
  panels show the pressure behavior of the corresponding transition
  temperatures $T_{C}^*$  and $T_{M}^*$ in the case of $\alpha_{2}^*=3.05$ 
  (b) and $\alpha_{2}^*=-3.05$ (c).  Results are shown distinctly for
  $\alpha_{1}^*=1$ (blue)  and $\alpha_{1}^*=-1$ (red).}  
\label{FIG4}
\end{figure}

The model for the metamagnetic transition corresponds to switch on the
parameter $\alpha_2^*$  in the free-energy model defined in
eq. (\ref{EQ5}). Notice that this parameter gives rise to an inversion
in the effective exchange constant that renders the $\cal{AF}$ stable
at low temperatures. It is worth mentioning that the importance of
magnetostriction in the occurrence of the $\cal{F} \rightarrow
\cal{AF}$ metamagnetic transition was first pointed out by
Kittel\cite{Kittel1960}. For the following calculation we  also take
$|\alpha_1^*|=1$ in order to favor discontinuous transitions.  In that
case,  both  order parameters, $x$ and $m$, may be different from zero
and the variation in the volume $w$ will depend on the sign of both
$\alpha_1^*$ and $\alpha_2^*$.  Standard numerical minimization of the
reduced Gibbs free-energy (\ref{EQ5}) predicts the occurrence of an
antiferromagnetic $\cal{AF}$-phase at low temperatures,  as it can be
seen in  the phase diagram shown in Fig. \ref{FIG4}(a). In this figure
we have plotted the behavior (in absence of external fields) of the
different transition temperatures as a function of the coupling
parameter $\alpha_2^*$ restricted to positive values for the sake of
clarity \footnote{The phase diagram in the region  corresponding to
  $\alpha_2^*<0$ is specularly similar with respect to
  $\alpha_2^*=0$.}. That is, the Curie temperature ${T^*}_{C}$
($\cal{P}$-$\cal{F}$), the Neel temperature ${T^*}_{N}$
($\cal{P}$-$\cal{AF}$) and the metamagnetic transition temperature
${T^*}_{M}$ ($\cal{AF}$-$\cal{F}$). The $\cal{AF}$-phase exists only
for values of the coupling parameter $\alpha_{2}^{*}>\alpha_{2c}^{*}$,
where $\alpha_{2c}^{*}$ satisfies\footnote{It can be easily derived by
  imposing that the  energy of both ${\cal F}$ and ${\cal AF}$ phases
  be equal at $T=0K$.}:
\begin{equation}
(\alpha_{2c}^*)^2-(\alpha_{1}^*)^2- 4 \Omega^* P
  (\alpha_{2c}^*-\alpha_{1}^*)-8 \alpha_0^* (1+H^*)=0
\label{EQ13}
\end{equation}
The temperature range
at which the $\cal{AF}$-phase exists increases with the coupling
strength $\alpha_{2}^*$.  There is a particular value,
$\alpha_{2t}^*$, at which the  three phases $\cal{P}$, $\cal{F}$ and
$\cal{AF}$ coexist. Thus, for
$\alpha_{2c}^*<\alpha_2^*<\alpha_{2t}^*$,  the model predicts two
consecutive phase transitions whereas for $\alpha_2^*>\alpha_{2t}^*$
the $\cal{F}$-phase disappears and the model exhibits an  unique
$\cal{P}$-to-$\cal{AF}$-phase transition at ${T^*}_{N}$.

\begin{figure}[ht]
\centering
\includegraphics[clip, scale=0.5]{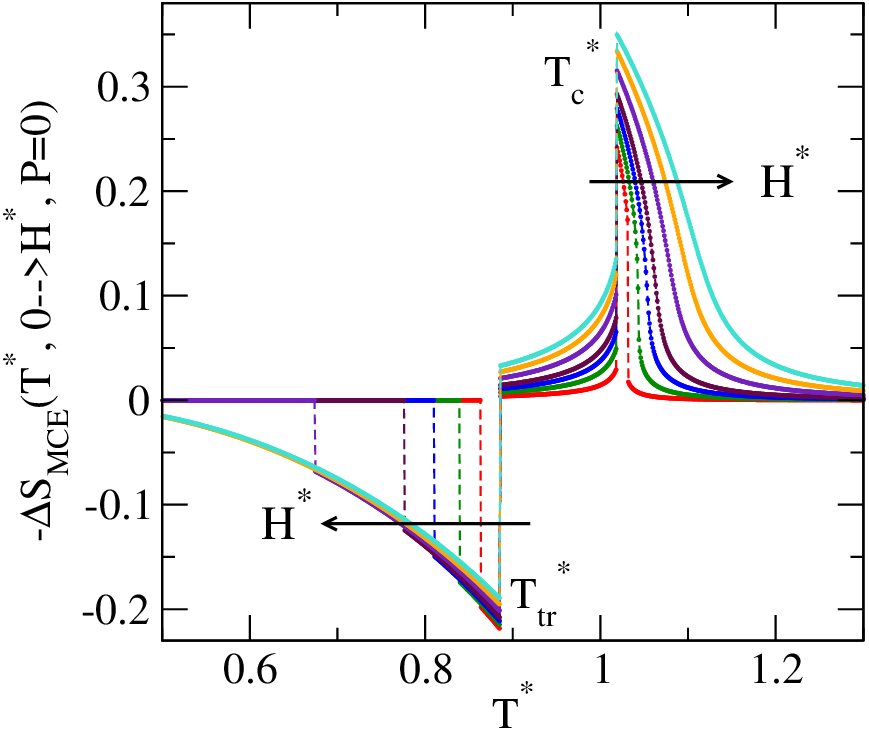}
\caption{(Color online) MCE for different values of the external
  magnetic field ranging from $H^*=0$ to $H^*=0.05$. $T_{C}^*$ and
  $T_{M}^*$ denote the ${\cal P}$-to-${\cal F}$ and ${\cal
    F}$-to-${\cal AF}$ transition temperatures respectively. As usual,
  the arrow denotes the direction of increasing $H^*$.}
\label{FIG5}
\end{figure}

Let us focus on the region of the
phase diagram where the metamagnetic transition exists  and take 
$|\alpha_2^*|=3.05$ (green line in Fig. \ref{FIG4}(a)).   In the lower
panels of Fig.\ref{FIG4} we show the corresponding behavior of both
transition temperatures, ${T^*}_{C}$ and ${T^*}_{M}$, with applied
external $P$ (b) for $\alpha_2^*=3.05$  and (c) $\alpha_2^*=-3.05$ 
respectively.  In both cases we explicitly distinguish between
$\alpha_1^*=1$ (blue) and $\alpha_1^*=-1$ (red).  One observes that
whereas for $\alpha_2^*= 3.05$  the application of $P$ tends to
suppress the ${\cal AF}$-phase rapidly, for $\alpha_2^*=- 3.05$  the
application of $P$ definitively renders the ${\cal AF}$-phase
favorable. Interestingly, the behavior displayed in Fig. \ref{FIG4}
(b) and (c) embodies whether the BCE is conventional (increasing
transition temperature with increasing $P$) or inverse (decreasing
transition temperature with increasing $P$). 
In Fig. \ref{FIG5} we show the MCE at different values of $H^*$
ranging from $H^*=0$ to $H^*=0.04$. The increasing stability of the
${\cal F}$ -phase is  reflected in the decrease of $T_{M}^*$ and the
simultaneous increase of $T_{C}^*$ with increasing $H^*$.   In
connexion with this, near the ${\cal P}$-to-${\cal F}$ transition
($T_{C}^*$), the MCE is conventional while  it is  inverse at lower
temperatures, around the ${\cal F}$-to-${\cal AF}$ transition
($T_{M}^*$). Moreover, the
conventional MCE peak increases with $H^*$ whereas the inverse MCE
peak decreases.  This apparent contradiction regarding the behavior of
the inverse MCE around the $\cal{F}$-to-$\cal{AF}$ transition  has to
do with the opposite effect that the application of $H^*$ has on  the
entropy of both  $\cal{F}$- and  $\cal{AF}$- phases.  In this sense,
the model predicts a sharp suppression of the $\cal {AF}$-phase that
hinders  a further increase of the entropy with increasing
$H^*$. To complete the discussion on the
MCE, it is worth mentioning that,  in adiabatic conditions, the system will
first warm up (at  high temperatures) and next cool down (at low temperatures) 
with the application of external $H^*$.
  

The effect of an external $P$ on the MCE  is displayed in the next
figure \ref{FIG6} where we have plotted the corresponding isothermal
entropy change at $P=0.015$ and selected values of the applied
magnetic field ranging from $H^*=0$ to $H^*=0.05$. Results have been
calculated for the different values $\alpha_{2}^*= \pm 3.05$ and
$\alpha_{1}^* = \pm 1$ considered previously. In general, the effect of
the secondary field is a temperature shift in the corresponding
caloric peak. As already mentioned, such displacement along the
temperature axis should be consistent with the behavior of the
transition temperatures displayed in  figures \ref{FIG4}(b) and
\ref{FIG4}(c).  Indeed,  an inspection of Fig. \ref{FIG6} reveals that
for $\alpha_2^*=3.05$ the effect of $P$  on the inverse MCE peak
(around $T_{M}^*$) is a shift to lower temperatures and a decay in the
response  whereas for $\alpha_2^*=-3.05$  the response gets enhanced
and shifted to higher temperatures.    Notice that for
$\alpha_{1}^*=-1$ the effect is dramatic  since  the application of
$P$ induces a further promotion  of the $\cal{F}$-phase.  Regarding to
the behavior of the conventional MCE around $T_{C}^*$, it has been
already discussed before. Similarly, the temperature shift in the
peaks follow the trends described in Fig. \ref{FIG4}(b) and (c).
\begin{figure}[ht]
\centering
\includegraphics[clip, scale=0.5]{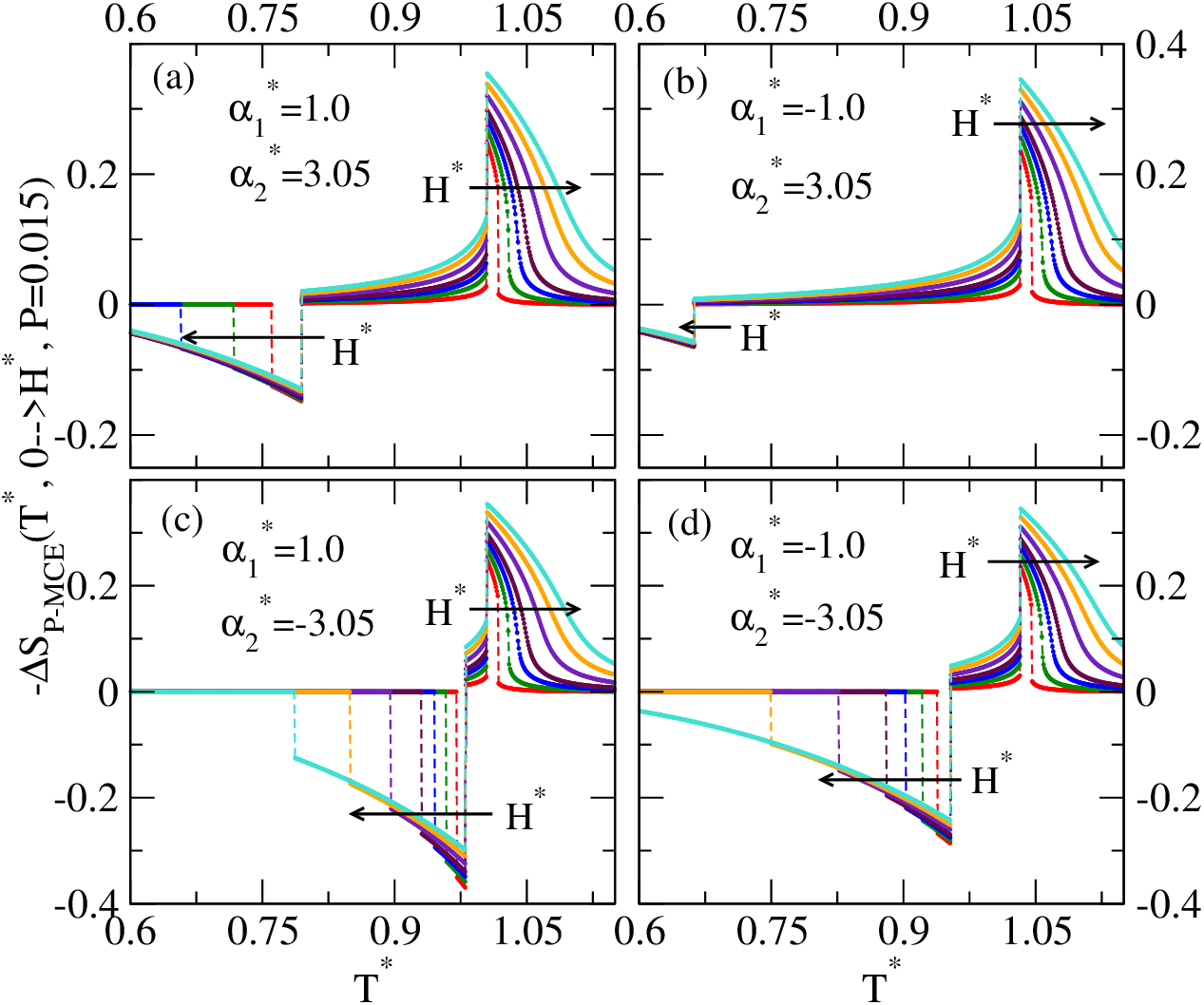}
\caption{(Color online) P-tune MCE for values of the applied external field
  ranging from  from $H^*=0$ to $H^*=0.05$ and $P=0.015$. Results are
  displayed distinctly for  $\alpha_{1}^*=\pm 1$ and  $\alpha_{2}^*=\pm 3.05$.}
\label{FIG6}
\end{figure}
\begin{figure}[ht]
\centering
\includegraphics[clip, scale=0.5]{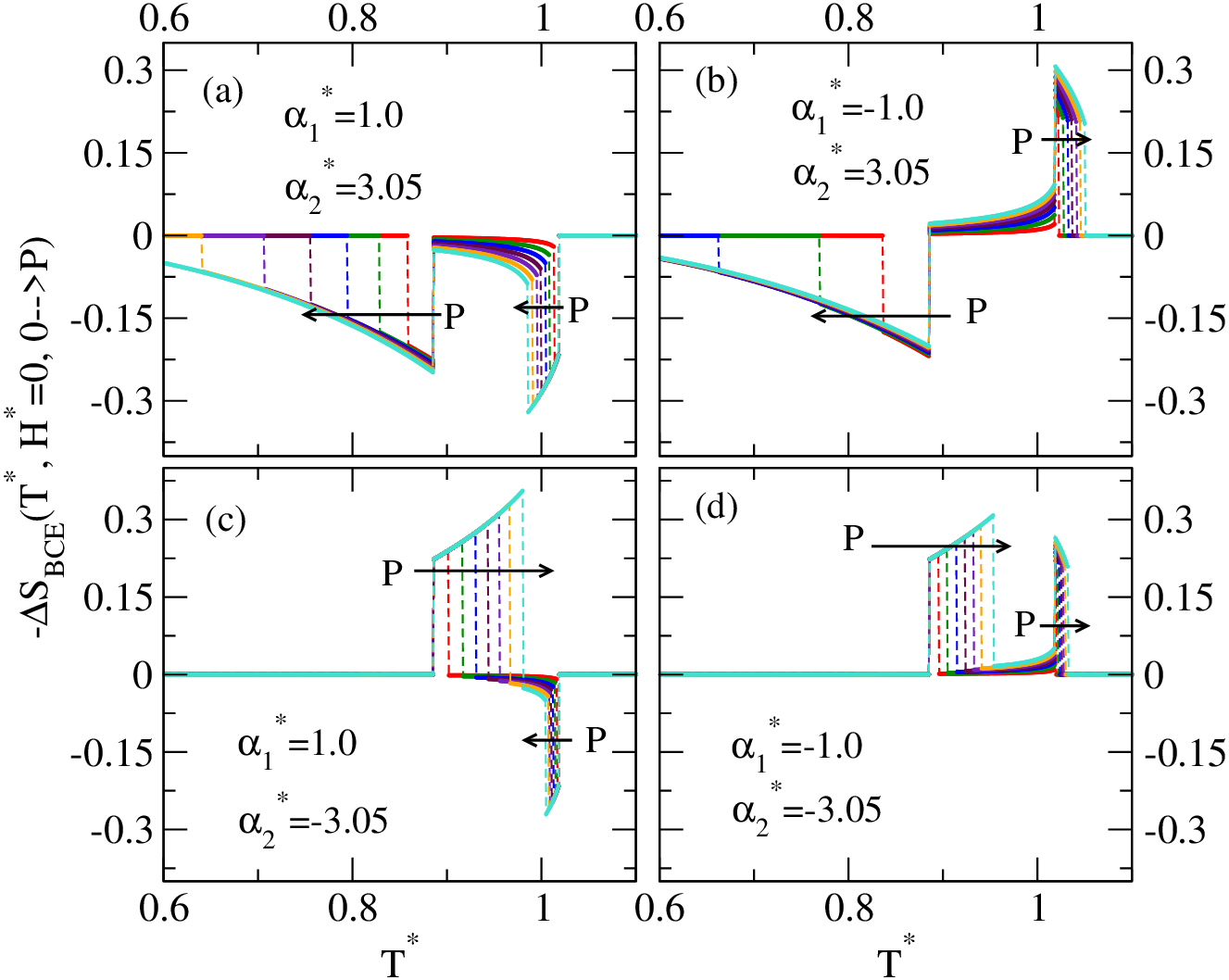}
\caption{(Color online) $\Delta S_{BCE}$ for increasing selected 
values of the applied pressure ranging from $P=0$ to
  $P=0.035$.  Results are displayed distinctly for $\alpha_{1}^*=\pm 1$  
and $\alpha_{2}^*= \pm 3.05$.}
\label{FIG7}
\end{figure}
\begin{figure}[ht]
\centering
\includegraphics[clip, scale=0.5]{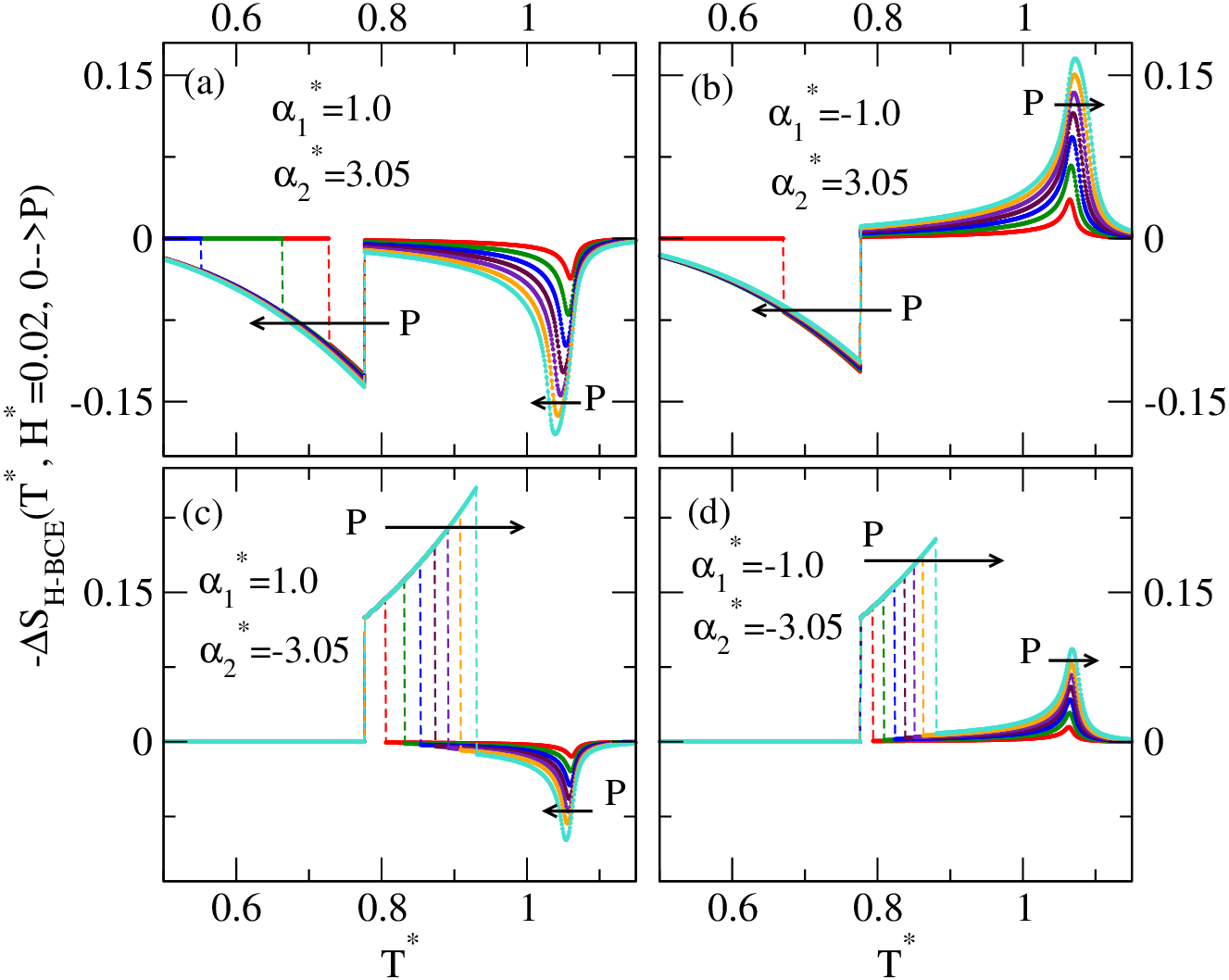}
\caption{(Color online) $\Delta S_{H-BCE}$ for increasing values of
  the applied pressure from $P=0$ to $P=0.035$ and $H^*=0.02$  
Results are displayed distinctly for $\alpha_{1}^*=\pm 1$  and $\alpha_{2}^*= \pm 3.05$.}
\label{FIG8}
\end{figure}

The results for the BCE behavior are shown in the next figure \ref{FIG7} 
for the same values of the coupling
parameters. The isothermal entropy change $\Delta S_{BCE}$
is displayed  for selected values of the applied pressure  ranging
from $P=0$ to $P=0.035$. Whereas the characteristics (whether it is
inverse or conventional) of the  high temperature peak around the
$\cal{P}$-$\cal{F}$ transition depends on the sign of $\alpha_{1}^*$,
$\alpha_{2}^*$ determines the characteristics  of the low
temperature peak around the  metamagnetic transition.
Thus, for $\alpha_{2}^*=3.05$ (panels  (a) and (b)) the  BCE 
around $T_{M}^*$ is inverse due to the suppression of the ${\cal
  AF}$-phase with increasing the applied pressure $P$ (Fig. \ref{FIG4}
(b)). Similarly, in the case of $\alpha_{2}^*=-3.05$ the
low-temperature BCE is conventional. Furthermore,  with increasing
$P$,  the BCE peak gets larger  for $\alpha_{1}^*=1$ and smaller for
$\alpha_{1}^*=-1$.  This is due to the fact that for positive
$\alpha_{1}^*$ the application of $P$ favors the disordered
$\cal{P}$-phase (in detriment of the $\cal{F}$-phase, with larger
volume)  while for negative values of $\alpha_{1}^*$ the ordered
$\cal{F}$-phase is promoted.  Concerning the second BCE peak around
$T_{C}^*$, it  is inverse for $\alpha_{1}^*=1$ and conventional for
$\alpha_{1}^*=-1$, consistently with the behavior of $T_{C}^*$ vs. $P$
shown in figures \ref{FIG4}(b) and \ref{FIG4}(c).  

To complete this section, in the different panels of figure \ref{FIG8}
we have included the effect of the secondary field ($H^*=0.02$) on the
previous BCE. A simple comparison with Fig. \ref{FIG7} reveals that
the effect of applying a magnetic field is to move away one peak from
the other and simultaneously to decrease the caloric response,
regardless of the model parameters. In summary, the application of
$H^*$  systematically reduces the response of the BCE and increases
the stability of the ($\cal{F}$)  phase.

\section{Relation to Experiments}
\label{Discussion}

In this section we analyze the previous theoretical results in
relation to the different caloric behaviors observed in magnetic and
metamagnetic materials for which experimental data is available. We
stress that discussion on the physical origin or mechanism  behind the
magnetoelastic coupling is out of the scope here. Rather, we shall
just require that the observation of the magnetic phase transition be
accompanied by some volume anomaly.  Below, we appraise our model
predictions, namely phase diagram and caloric responses, by comparing
them with experiments in the case of two potential magnetic
refrigerant materials, $La_{(1-x)}Ca_{x} MnO_{3}$ and $FeRh$.
Qualitative information regarding general aspects such as whether the
caloric effect is conventional or inverse and  the behavior (i.e. the
temperature shift) of the caloric peak under the application of a
secondary field can be inferred directly from the phase diagram. Even
so, the  maximum value of the caloric response, either on $\Delta
S_{T}$ or $\Delta T_{a}$ might depend on other aspects or
contributions not described properly (or not described) in the model.

\subsection{The $La_{(1-x)}Ca_{x} MnO_{3}$ CMR system}

\begin{figure}[ht]
\centering
\includegraphics[clip, scale=0.5]{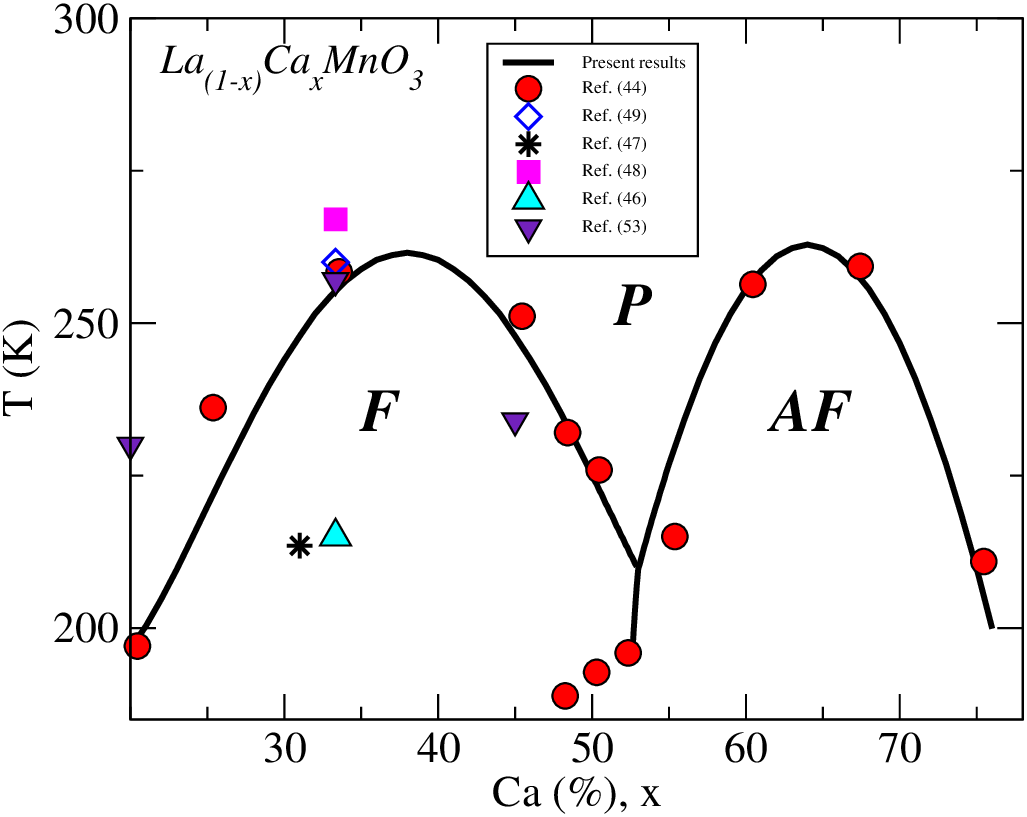}
\caption{(Color online) Phase Diagram for  $La_{1-x}Ca_{x}MnO_{3}$ as
  a function of the Ca content ($x$). Continuous line denotes the
  present numerical results obtained by  assuming a quadratic dependence
  of the coupling parameters   $\alpha_{1}^*$ and $\alpha_{2}^*$ with
  doping $x$. Points correspond to experimental data from different authors
  indicated in the inset.}
\label{FIG9}
\end{figure}

Few years ago, very much attention was given to the study of the
$La_{(1-x)}Ca_{x} MnO_{3}$ perovskite because of the unexpected large
magnetoresistance observed at low temperatures \cite{Jim1994}.  As a
function of temperature and doping ($x$), this material shows
different magnetic transitions \cite{Schiffer1995}. When lowering the
temperature, it exhibits a  $\cal{P} \rightarrow \cal{F}$ transition
for $x \leq 0.50$ and  a  $\cal{P} \rightarrow \cal{AF}$ transition
for $x \geq 0.5$.  From the point of view of the present model, such a
different  magnetic behavior can be taken into account by  recalling
that both coupling coefficients, $\alpha_{1}^*$ and $\alpha_{2}^*$, are
composition dependent.  In figure \ref{FIG9} we present the results
obtained for the transition temperature assuming a quadratic
dependence with doping for both coefficients. The present numerical
results  (denoted by a continuous line) have been  obtained by taking
$\alpha_{1}^*= 24.22 (x-0.38)^2-2.5$ and $\alpha_{2}^*=-22.45
(x-0.64)^2 +3.8$. The corresponding estimation for the exchange
constant is  $zJ=16.6$meV, close to the values  (6.6-10.7) meV
reported in the literature\cite{Nicastro2002}. The model results are
compared with available experimental data  taken from different
authors, as indicated in the inset.  We might conclude that  the
agreement is remarkable.  The unusual deviation around $x \sim 30\%$
is attributed\cite{DeTeresa1996,sun1,sun2,Zhang1996}  to differences
in the method used in preparing the sample.  Interestingly, close to
$x \sim 0.50$ the ground state changes from $\cal{F}$ to $\cal {AF}$
although the direct $\cal{F}  \rightarrow \cal{AF}$ metamagnetic
transition (if possible) will be restricted to a very  narrow interval
of values of $x$. Actually, metamagnetic transitions in
$La_{(1-x)}Ca_{x} MnO_{3}$ have only been reported under the
application of (low) external magnetic fields\cite{Ulyanov2008}.

\begin{figure}[ht]
\centering 
\includegraphics[clip, scale=0.5]{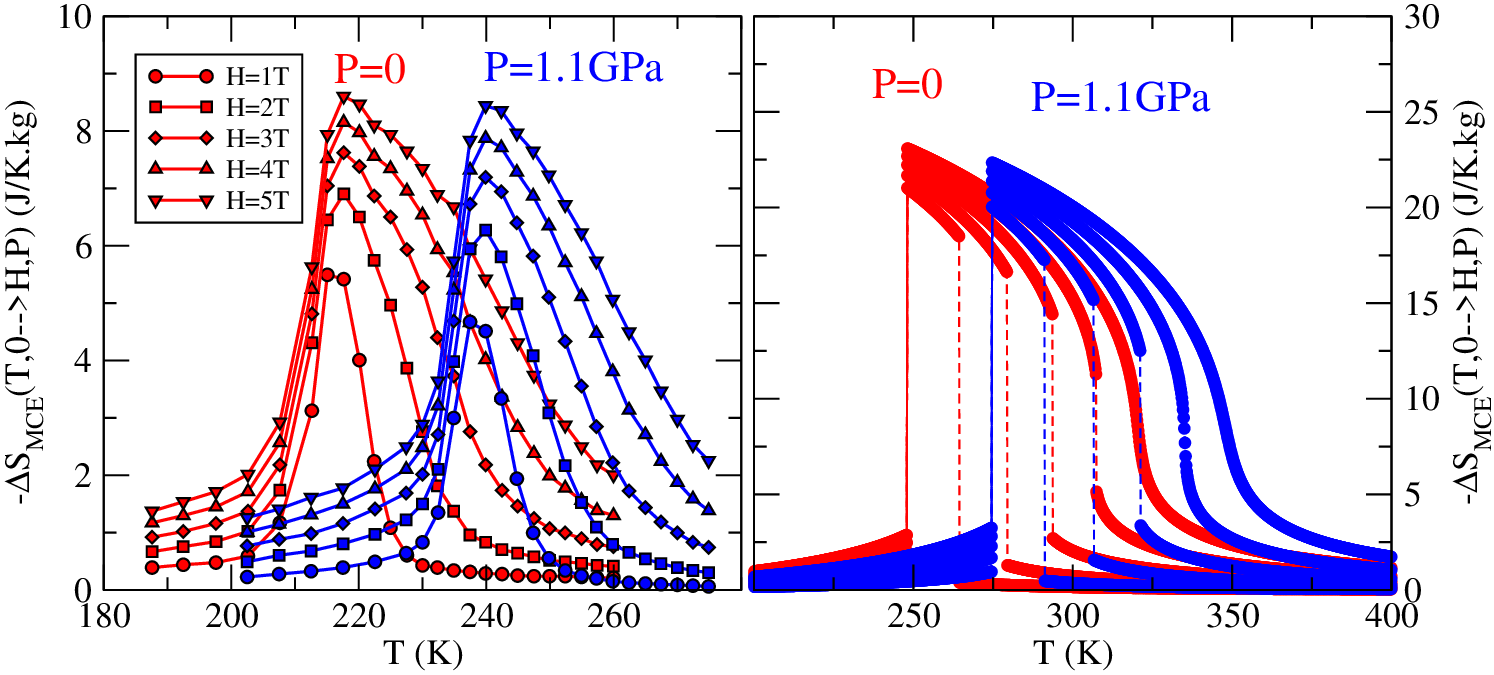}
\caption{(Color online). Effect of the pressure on the $\Delta S_{MCE}$
  in $La_{0.69}Ca_{0.31}MnO_{3}$. (a) corresponds to experimental data
  taken from Ref.(47) and (b) displays the present numerical
  results for the same values of applied fields and pressures.}
\label{FIG10}
\end{figure}

Moreover, perovskite manganites show a strong spin-lattice coupling
\cite{Guo1997}.  This makes the study  of pressure effects on their
magnetic behavior of potential interest.  In fig. \ref{FIG10} we show
the effect of pressure on the MCE in $La_{0.69}Ca_{0.31} MnO_{3}$.
Panel (a) displays the experimental data\cite{sun1} for different
values of the applied magnetic field and for two  values of the
applied pressure, $P\eqsim 0$ (ambient pressure) and $P= 1.1$GPa. In panel
(b) we have plotted the present numerical results obtained for
$\alpha_1^*= -2.38$ and $\alpha_2^*=1.36$ and for the same values of
the fields as in panel (a). We obtain an estimation for the volume of
the unit cell of $\Omega \eqsim 29(\AA)^3$,  one half of the
experimental value\cite{Radaelli1995,Nicastro2002} ($\sim 60(\AA)^3$)
but with the right order of magnitude.   Before continuing with the
discussion of fig. \ref{FIG10},  let us point out that for these
values of the coupling coefficients, $|\alpha_{2}^*|<|\alpha_{2c}^*|$
(defined in eq. (\ref{EQ13})). The model predicts an unique $\cal{P}
\rightarrow \cal{F}$ transition  at a temperature that increases with
both applied magnetic field and pressure. In this situation, the
parameter $\alpha_2^*$ is irrelevant and consequently the description
can be done by means of the simplified model  defined in section
\ref{Ferro}, with $\alpha_1^*<0$.  Indeed, our results preview  that
both MCE (fig. \ref{FIG2} (a) and (b)) and BCE (fig. \ref{FIG3} (b)
and (d)) are conventional with behavior under external fields  given
in figures \ref{FIG2}(d) and \ref{FIG3}(f).   We now return to figure
\ref{FIG10}. A simple inspection reveals that in this material the
main effect of pressure on the MCE is a simple  shift of the whole
response (almost unaltered) to higher temperatures consistently with
the increasing stability of the $\cal{F}$-phase (with lower volume)
with $P$.  In conclusion, the  model is able to reproduce the general
experimental  trends. Nevertheless, the amount of  $\Delta S_{MCE}$
even though has the  right order of magnitude is underestimate by a
factor two (roughly).  We attribute this to other entropy
contributions, mainly electronic, not considered in the present model.  

Very briefly we would like to mention that similar behavior is observed in
$La(Fe_{x}Si_{1-x})_{13}$ -type compounds. The field-induced
first-order $\cal P$-to-$\cal F$ phase transition when lowering the
temperature, is accompanied by a significant isotropic expansion of
the volume and the  application of an external $P$ reduces the Curie
temperature\cite{Lyubina2008}. Again, the description of the general
trends  can be done by  the simplified model (\ref{EQ9}) but now with
$\alpha_1^* >0$.  It has been reported that the MCE is
conventional\cite{Fujita12003,Hu1,Manosa2011} whereas the BCE is
inverse\cite{Manosa2011}. Indeed, results shown in figures
\ref{FIG2}(a)  and \ref{FIG3}(a) are consistent with such experimental
behavior.  Additionally, the tunning of the MCE by an external
pressure shifts the whole caloric effect towards lower temperatures
\cite{Lyubina2008} and the BCE exhibits a negative adiabatic
temperature change \cite{Manosa2011}. These trends are reproduced  in
figures \ref{FIG2}(c) and \ref{FIG3}(c).

\subsection{The FeRh metamagnetic  alloy}

The $B2$ ordered near-equiatomic $Fe_{1-x}Rh_{x}$ alloy displays a
metamagnetic transition  from an $\cal{AF}$ ground state to a
$\cal{F}$-phase with increasing temperature.  It occurs  around $T
\sim 320K$ and it is accompanied by a $1\%$ volume increasing in the
unit cell that preserves the cubic symmetry
\cite{Kouvel1966,Nikitin1992,Gruner2003,Barua2013} . This singular
transition is strongly concentration  dependent \cite{Staunton2014}
and it is only present for a very narrow  range of the composition
($0.48 \leq x \leq 0.52$)\cite{Barua2013}. Additionally, it also
depends on heat treatment \cite{Annaorazov1996}, configurational
ordering \cite{Staunton2014,Sandratskii2011} and external fields.  Of
special interest is the study of pressure effects on the magnetic
behavior
\cite{Ponyatovskii1968,Heeger1970,Dubovka1974,Vinokurova1976,Gruner2005}. The
$\cal{F}$-phase, in between the $\cal{P}$-phase (at high temperatures)
and the $\cal{AF}$-phase (at low temperatures), exists only for values
of the applied pressure below $\sim 6$ GPa (tricritical pressure). For
higher pressures the metamagnetic transition disappears and the
$\cal{AF}$-phase transforms into the $\cal{P}$-phase directly. In the
next figure \ref{FIG11} we show the $P$-$T$ phase diagram for the
nominally equiatomic $Fe_{(1-x)}Rh_x$ ($x \eqsim 0.5$). Continuous
(blue) lines denote the present results for $\alpha_1^*= 1$ and
$\alpha_2^*=-3.0$ (see fig. \ref{FIG4}) whereas points correspond
to experimental data taken from different authors 
indicated in the inset. The fitting to the experimental data renders
the following estimations $zJ=5.66$meV and $\Omega=36(\AA)^3$,
comparable to recently reported values for the exchange constant
\cite{Kudrnovsky2015} and the lattice parameter \cite{Barua2013}
respectively.  The misfit  between theory and experiments ($\sim 10
\%$ around the tricritical point) is partially due to the more attention given
in the fitting procedure at the behavior close to P=0.   In spite of
this, we conclude that the agreement is satisfactory.

\begin{figure}[ht]
\centering
\includegraphics[clip, scale=0.5]{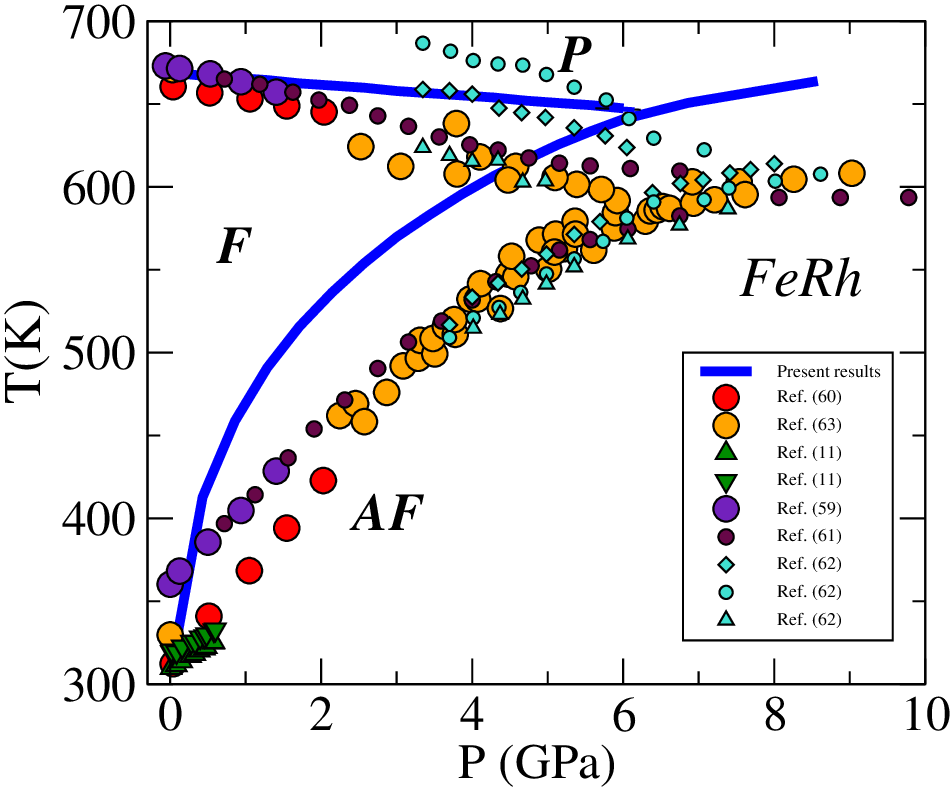}
\caption{(Color online)  $P$-$T$ phase diagram for the equiatomic
  $FeRh$ alloy.  The present numerical results (blue line) are
  compared with available experimental data taken from different
  authors indicated in the inset.}
\label{FIG11}
\end{figure}
\begin{figure}[ht]
\centering
\includegraphics[clip, scale=0.5]{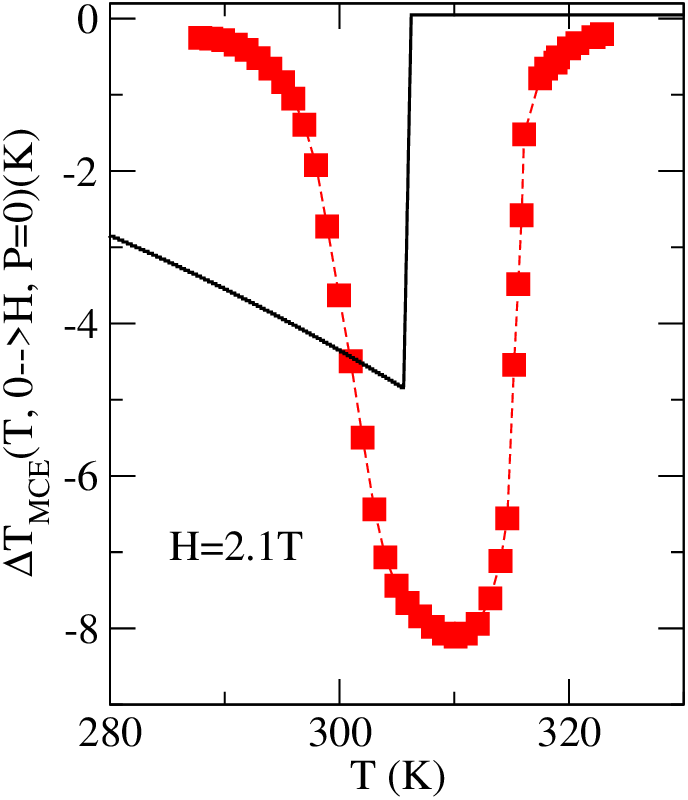}
\caption{(Color online) Adiabatic temperature change in the MCE at
  H=2.1T in FeRh alloy. Symbols correspond to experimental data from
  Ref. (27) whereas the continuous line indicate the
  present results.}
\label{FIG12}
\end{figure}

Concerning the caloric behavior in $FeRh$ metamagnetic alloy  near the
${\cal F}$-${\cal AF}$ transition, experiments show that the MCE is
inverse \cite{Annaorazov1996,Stern2014} while  the BCE is conventional
\cite{Stern2014}. Under the application of an external pressure this
transition temperature increases while the Curie temperature decreases
\cite{Heeger1970}.  This scenario is reproduced  by the present
theoretical predictions shown in figures \ref{FIG6} (c) and
\ref{FIG7}(c). In Figure \ref{FIG12} we show the cooling by the adiabatic
magnetization (as expected for an inverse MCE) as  observed near the
metamagnetic transition in $FeRh$. Experiments \cite{Annaorazov1996}
are denoted by symbols whereas the continuous line corresponds to our
results. These last have been obtained from the entropy curves by
requiring that $S(T_f, H=2.1T,P=0)=S(T,0,0)$ and  using
$\Theta_D=400K$. Experimentally, the maximum cooling at $H=2.1T$ is of
$\Delta T_{exp}=-8 K$ whereas we obtain $\Delta T= -4.9K$.  

Unfortunately the present model predicts a value  for the entropy
change (either in the MCE or BCE) in $FeRh$  one order of magnitude
below the experimental value\cite{Stern2014,Annaorazov1996}  ($|\Delta
S_{exp}|= 12J K^{-1}Kg.^{-1}$). Principally this is due to the subtle
balance between the different entropy contributions \cite{Cooke2012}
$\Delta S_{mag}$ (magnetic), $\Delta S_{v}$ (lattice) and  $\Delta
S_{elec}$ (electronic)  and the crucial  role played by this last in
ensuring a large enough value for the total amount of the entropy
change. Although still under debate,  it is  accepted that in $FeRh$
the $\cal{AF} \rightarrow \cal{F}$ metamagnetic transition is driven
by an excess of electronic and magnetic entropy while the lattice
opposes to the transition.  Roughly speaking  $\Delta S_{v}  \eqsim
-70 \%\Delta S_{mag}$  and  $\Delta S_{elec}$ represents a $40\%$ of
the  total $\Delta S$. This balance makes our model- that does not
consider the electronic contribution- unqualified to obtain  a
reasonable value of the entropy change in this material. Nevertheless
it predicts a quite acceptable  value for the adiabatical temperature
change due to the satisfactory description of the $P$-$T$ phase diagram. 
In this regard, it should be mentioned that completely
adiabatic conditions are very difficult to achieve
experimentally. Finally let us noting  a recent
study\cite{Barua2013} aimed at finding out magnetoestructural trends
in FeRh-based alloys. In particular, the behavior of the transition
temperature as a function of the valence electron per atom seems to
confirm the importance of the electronic effects on the
transition. Also,  these results seem to indicate that magnetovolumic
effects are not essential for the transition although they are crucial
in stabilizing the low temperature $\cal{AF}$- phase.

\section{Conclusions}
\label{conclusions}

We present a mean-field Landau-based  model for phase transitions that
captures the main ingredients necessary to reproduce the phase diagram
and the general trends of the experimental caloric behavior observed
in magnetoelastic materials in response to the application of
external fields, either magnetic or/and hydrostatical pressure. In
particular, we have applied the results to $LaCaMnO_3$ perovskite and
to $FeRh$ metamagnetic alloy. Such materials are very different but
have the common feature of undergoing a magnetic phase transition
accompanied by magnetoelastic effects. This is enough for the  model
to be able to reproduce both phase diagrams to a very good level of
agreement with the experiments.  The main limitation of the model is
to predict the correct order of magnitude  for the entropy change at
the metamagnetic transition. Apparently, this is due to the fact that it
includes the magnetic degrees of freedom only disregarding the role of
the electronic contribution that in this material turns out to be very
important.  Concerning the lattice contribution, it plays the role of
a thermal bath for the adiabatic caloric process. In this sense,
Gruner {\it et al.}, by performing  Monte Carlo simulations of a
spin-based model extended to include magnetovolumic effects, were able
to obtain a value for the entropy change within the range of the
experimental results \cite{Gruner2003}. This could be indicative of
the importance of fluctuations in the occurrence of metamagnetic
transitions. Additionally, coupling coefficients  
can be evaluated from first principle calculation thus providing 
an estimation independent on the model.



\begin{acknowledgments} 
This work has received financial support from CICyT (Spain), Project
No. MAT2013-40590-P.  One of us (E.M.) thanks the Spanish Ministery of
Eduaction, Culture and Sports for the fellowship for collaboration
with the Dept. d'Estructura i Constituents de la Materia (UB) during
his last year of undergraduate student in Physics.
\end{acknowledgments}
\bibliography{bibliography}

\end{document}